\documentclass[aps,eqsecnum,preprint,preprintnumbers,12pt,amsfonts]{revtex4}
\usepackage{epsfig,psfrag}

\def\bea{\begin{eqnarray}}
\def\eea{\end{eqnarray}}
\def\nn{\nonumber}

\newcommand\eqn[1]{(\ref{#1})} 

\begin{document}

\title{Large-order Perturbation Theory and de Sitter/Anti de Sitter Effective Actions}
\author{Ashok Das} \email{das@pas.rochester.edu}
\affiliation{Department of Physics and Astronomy, University of Rochester, Rochester, NY 14627-0171} \affiliation{Saha Institute of Nuclear Physics, 1/AF Bidhannagar, Calcutta 700064, INDIA}
\author{Gerald V. Dunne}\email{dunne@phys.uconn.edu}
\affiliation{Department of Physics, University of Connecticut, Storrs, CT 06269-3046}

\begin{abstract}
We analyze the large-order behavior of the perturbative weak-field expansion of the effective Lagrangian density of a massive scalar in de Sitter and anti de Sitter space, and show that this perturbative information is not sufficient to describe the non-perturbative behavior of these theories, in contrast to the analogous situation for the Euler-Heisenberg effective Lagrangian density for charged scalars in constant  electric and magnetic background fields. For example, in even dimensional de Sitter space there is particle production, but the effective Lagrangian density is nevertheless real, even though its weak-field expansion is a divergent non-alternating series whose formal imaginary part corresponds to the correct particle production rate. This apparent puzzle is resolved by considering the full 
non-perturbative structure of the relevant Feynman propagators, and cannot be resolved solely from the perturbative expansion.
\end{abstract}

\maketitle

\section{Introduction}
\label{intro-sec}
A fundamental result in the study of effective actions in 
gauge theories is the Euler-Heisenberg QED effective action arising from 
a single spinor (or charged scalar) loop in the presence of a constant 
field strength $F_{\mu\nu}$ \cite{he,weisskopf,schwinger,dunnekogan}. This result serves as the
starting point for many calculations of vacuum polarization effects both
in QED and QCD. The effective action can be found non-perturbatively in
this case because in such a constant background field, the Dirac (or Klein-Gordon) operator has a
simple spectrum. A weak-field expansion of the non-perturbative answer can be
identified term-by-term with the  perturbative diagrammatic
expansion of the effective action. For example, for a constant electric
or magnetic field the weak-field expansion of the effective action consists of a perturbative 
series with coefficients that grow factorially  in magnitude, a behavior typical of a
wide range of perturbation theory problems in physics \cite{leguillou}. For a constant $E$ field, the perturbative series is a non-alternating divergent series, whose non-perturbative imaginary part is associated (via a Borel dispersion relation) with the instability of the 
vacuum in an electric field background \cite{popov,chadha}.

A natural gravitational analog of the constant electromagnetic field case is a theory in a manifold with a constant curvature \cite{candelas,dowker}. de Sitter (dS) and anti de Sitter (AdS) spaces provide such a background and in this paper we apply such a large-order perturbation theory analysis to the effective action for a massive scalar field (the analysis for spinors is very similar) in a de Sitter or anti de Sitter background.
Since these backgrounds have constant curvature $R$, we expect the gravitational effective action 
to have a weak-field expansion of the form \cite{donoghue,buchbinder,eric}
\begin{eqnarray}
S=\int d^{d+1}x \sqrt{g}\left(a_1 R+a_2 R^2+a_3 R^3+\dots \right)
\label{grav}
\end{eqnarray}
where the $a_n$ are (dimensionful) expansion coefficients. As in the case of the Euler-Heisenberg action, the de Sitter and anti de Sitter backgrounds are sufficiently simple that the effective actions can be computed in closed-form without resort  to perturbation theory in $R$ \cite{tagirov,candelas,dowker,isham,allen,lapedes,mottola,allen2,burgess,camporesi,kamela,dowker-sphere,spradlin,bousso}. Therefore, one can carry out a complete large order perturbative analysis of the associated weak-field expansion, and study its Borel properties and possible non-perturbative imaginary parts, just as in the QED case in \cite{popov,chadha}. This provides information about the scalar loop contribution to the low energy graviton affective action \cite{donoghue,buchbinder,eric}. In odd dimensional space-time, for both AdS and dS, one finds that the effective action has a convergent weak-field expansion, and correspondingly there is no non-perturbative imaginary part, consistent with the absence of particle production in odd dimensions \cite{lapedes,mottola,allen2,bousso}. In even dimensions, the AdS case is Borel summable and real, once again consistent with the absence of particle production in AdS space. However, for even dimensional dS space the weak-field pertrubative series is a non-alternating divergent series, for which a formal Borel analysis along the lines of \cite{popov,chadha} leads to a non-perturbative imaginary part coinciding with the particle production rate computed through a Bogoliubov transformation analysis \cite{lapedes,mottola,allen2,bousso}. But the even dimensional dS effective Lagrangian density is manifestly real, so this shows that the large-order perturbative behavior alone is not sufficient to capture the correct non-perturbative physics. In this paper we explain and resolve these issues.

The paper is organized as follows. In Section {\bf II} we recapitulate briefly the well known Euler-Heisenberg results for a charged scalar field interacting with a constant background magnetic/electric field. We express the effective Lagrangian densities for these cases in a way that can be compared with the gravitational case described later. We recall the Borel summation analysis in some detail since it is quite relevant for the gravitational case. In Section {\bf III}, we derive the effective Lagrangian densities for a scalar field in a de Sitter/anti de Sitter background in dimension $d$, through the method of the coincidence limit of the Feynman propagator. We carry out a weak curvature expansion of the effective Lagrangian density and analyze the behavior of the large order terms in this perturbative series. We show that the effective action obtained through Borel summation is real in anti de Sitter space as well as in odd dimensional de Sitter spaces. The Borel analysis of the even dimensional dS case is more subtle, and raises an interesting puzzle concerning the connection between perturbative and non-perturbative physics in gravitational theories. We present a resolution of this puzzle in Section {\bf IV}, and conclude with a brief summary in Section {\bf V}. In the appendix, we collect some formulas on the multiple gamma functions that are used in the text.

\section{Scalar Effective Action in Electromagnetic Backgrounds}
\label{gauge-sec}

In this section, we briefly review the Borel summation analysis \cite{popov,chadha} of the Euler-Heisenberg effective action for charged particles in a constant electromagnetic background field \cite{he,weisskopf,schwinger,dunnekogan}.
First, consider scalar particles in a constant magnetic field of strength $B$.  The effective Lagrangian density for the theory  can be expressed in terms of a proper-time integral of the form (we suppress factors of electric charge):
\bea
{\mathcal L}(B)=\left(\frac{B}{4\pi}\right)^2 \int_0^\infty \frac{ds}{s^2}  \, e^{-\frac{m^2}{B}\, s} \left[\frac{1}{\sinh(s)} -\frac{1}{s}+\frac{s}{6}\right]   \quad ,
\label{bpt}
\eea
where $m$ denotes the mass of the charged particle and $B$ the magnitude of the magnetic field.
The first subtraction is the free-field subtraction and the last term corresponds to charge renormalization.
This integral is easily seen to  converge for all positive $\frac{B}{m^2}$, and  
an asymptotic expansion of the integral in the weak field limit yields the perturbative expansion : 
\bea
{\mathcal L}(B)\sim \left(\frac{m^2}{4\pi}\right)^2 \sum_{n=0}^\infty \frac{\bar{\mathcal B}_{2n+4}}{(2n+2)(2n+3)(2n+4)}\,\left(\frac{2B}{m^2}\right)^{2n+4} \quad .
\label{bexp}
\eea
Here the modified Bernoulli numbers are defined as: $\bar{\mathcal B}_{2n}\equiv \left(2^{1-2n}-1\right){\mathcal B}_{2n}$. Note that only even powers of $B$ appear, a reflection of Furry's theorem.
The perturbative coefficients in \eqn{bexp} alternate in sign and grow factorially  in magnitude \cite{as}
\bea
a_n^{(B)}&\equiv& \frac{2^{2n+4}\, \bar{\mathcal B}_{2n+4}}{(2n+2)(2n+3)(2n+4)}\nn\\
&\sim& 2(-1)^{n-1} \frac{\Gamma[2n+2]}{\pi^{2n+4}} \sum_{k=1}^\infty \left(\frac{2}{(2 k)^{2n+4}}-\frac{1}{k^{2n+4}}\right)\quad .
\label{b-growth}
\eea
Thus, the perturbative weak-field expansion \eqn{bexp} is an alternating divergent series, and is Borel summable. Indeed, using the basic Borel summation relation \cite{zinn,thooft}
\bea
\sum_{n=0}^\infty (-1)^n \Gamma(\beta n+\gamma) \, g^n  
\sim \frac{1}{\beta}\int_0^\infty \frac{ds}{s}\, \left(\frac{1}{1+s}\right) \,\left(\frac{s}{g}\right)^{\gamma/\beta}\,\exp\left[-\left(s/g\right)^{1/\beta}\right] \quad ,
\label{borel}
\eea
together with the large-order growth \eqn{b-growth} of the perturbative expansion coefficients, 
one can show that the proper-time expression \eqn{bpt} is indeed the Borel sum of the divergent series \eqn{bexp} 
[Recall the expansion $\frac{1}{\sinh(s)}=\frac{1}{s}-\frac{s}{6}+2s^3\sum_{k=1}^\infty \frac{(-1)^{k-1}}{(k\pi)^2\left(s^2+(k \pi)^2\right)}$]. This is, of course, just a self-consistency check, because the weak-field expansion \eqn{bexp} was obtained from the closed-form nonperturbative expression \eqn{bpt} in the first place. For later comparison with the gravitational case, we list here an alternative integral representation of ${\mathcal L}(B)$, in terms of the digamma function, $ \psi (z) = \frac{d \ln \Gamma(z)}{dz}$, [which can in turn be written in terms of the multiple gamma function $\Gamma_2$ \cite{barnes}, using the result \eqn{magic} in the appendix]:
\bea
{\mathcal L}(B)
&=&\left(\frac{B}{2\pi}\right)^2 \left\{ \int_0^{\frac{m^2}{2B}}x\, \psi\left(\frac{1}{2}+x\right) dx-\frac{m^2}{2B}\ln \Gamma\left(\frac{1}{2}+\frac{m^2}{2B}\right)-\frac{3}{4}\left(\frac{m^2}{2B}\right)^2+\frac{m^2}{4 B} \ln(2\pi) \right.\nn\\
&&\left. +\frac{\ln 2-1}{24}+\frac{1}{2}\zeta^\prime(-1)-\left(\frac{1}{24}-\frac{1}{2}\left(\frac{m^2}{2B}\right)^2\right)\ln \left(\frac{m^2}{2B}\right)\right\} \quad .
\label{multiplegamma}
\eea

For a constant {\it electric} field background instead, the only change in the perturbative series expansion \eqn{bexp} is the replacement $B^2\to -E^2$, which results from the sole Lorentz invariant quantity $(B^{2}-E^{2})$, when only one or the other field is present. Thus the perturbative weak-field expansion in this case is
\bea
{\mathcal L}(E)
\sim \left(\frac{m^2}{4\pi}\right)^2 \sum_{n=0}^\infty \frac{(-1)^n \bar{\mathcal B}_{2n+4}}{(2n+2)(2n+3)(2n+4)}\,\left(\frac{2E}{m^2}\right)^{2n+4}\quad .
\label{eexp}
\eea
The difference compared to the magnetic case \eqn{bexp} is that the series \eqn{eexp} is {\it non-alternating}: 
\bea
a_n^{(E)}=(-1)^n a_n^{(B)}\quad .
\label{ebalt}
\eea
Thus, the weak-field expansion \eqn{eexp} is not Borel summable. Nevertheless, the conventional {\it formal} Borel prescription \cite{vainshtein,benderwu,zinn,thooft} leads to an imaginary part for a divergent non-alternating series of this form :
\bea
{\rm Im}\left(  \sum_{n=0}^\infty \Gamma(\beta n+\gamma) \, g^n\right) &\sim &
{\rm Im}\left(\frac{1}{\beta}\int_0^\infty \frac{ds}{s}\, \left(\frac{1}{1-s}\right) \,\left(\frac{s}{g}\right)^{\gamma/\beta}\,\exp\left[-\left(s/g\right)^{1/\beta}\right]\right)\nn\\
&\sim& \frac{\pi}{\beta}\left(\frac{1}{g}\right)^{\gamma/\beta}\,\exp\left[-\left(1/g\right)^{1/\beta}\right]\quad .
\label{imag-borel}
\eea
Applying this Borel formula, together with the large-order growth of the perturbative coefficients in \eqn{b-growth}, one derives the well-known result for the imaginary part of the effective Lagrangian density:
\bea
{\rm Im}\left({\mathcal L}(E)\right) \sim \frac{E^2}{16\pi^3}\sum_{k=1}^\infty \frac{(-1)^{k-1}}{k^2} \,
\exp\left[-\frac{m^2 \pi k}{E}\right]\quad .
\label{e-imag}
\eea
These formal steps amount to assuming no further poles or cuts in the Borel plane beyond the simple pole at $s=1$ in \eqn{imag-borel}, for which a principal parts prescription is used for the calculation of the imaginary part. The {\it a posteriori} justification for these manipulations is that the final result agrees with that obtained by analytically continuing the Hurwitz zeta function or multiple gamma function expression for ${\mathcal L}(B)$ under $B\to iE$, and gives the correct non-perturbative proper-time expression:
\bea
{\mathcal L}(E)=\left(\frac{E}{4\pi}\right)^2 \int_0^\infty \frac{ds}{s^2}\, e^{-\frac{m^2}{E}\, s} \left[\frac{1}{\sin(s)} -\frac{1}{s}-\frac{s}{6}\right] \quad .
\label{ept}
\eea
This is essentially the argument of \cite{popov,chadha}, adapted from spinor QED to scalar QED. It shows that from a knowledge of the large order divergence behavior \eqn{b-growth} of the perturbative expansion, we can deduce non-perturbative information about the imaginary part of the effective Lagrangian density, {\it under the assumption that no other Borel poles or cuts contribute}. In certain cases one can even extend this analysis to inhomogeneous background fields, with the derivative expansion capturing non-perturbative information in its divergence \cite{dhborel}. These results provide a working illustration of Dyson's formal physical argument \cite{dyson} concerning the divergence of QED perturbation theory: one can argue that the weak-field expansions \eqn{bexp} and \eqn{eexp} could not be convergent, because if they were, they could not capture the genuine non-perturbative effect of pair production. This idea of connecting non-perturbative physics with the large-order behavior of perturbation theory goes back to the fundamental quantum mechanical analyses of Vainshtein in $\lambda x^3$ theory \cite{vainshtein}, and Bender and Wu in $\lambda x^4$ theory \cite{benderwu}, as well as the field theory arguments of Lipatov \cite{lipatov}.

The main point of this paper is to show that while the gravitational cases of de Sitter and anti de Sitter backgrounds are very similar to these constant electromagnetic backgrounds, they have the important difference that this naive Borel approach is not sufficient to deduce the true non-perturbative behavior. In particular,  genuine non-perturbative information concerning the Feynman propagators is required in order to compute the correct effective Lagrangian density. A simplistic application of Borel formulas like \eqn{imag-borel} is not enough to bridge the gap between the perturbative expansion and the non-perturbative structure of the theory.

\section{Scalar Effective Action in Gravitational Backgrounds}
\label{gravity-sec}

As noted by many authors, a natural generalization of the constant field strength electromagnetic backgrounds to gravity is that of constant curvature gravitational backgrounds \cite{candelas,dowker}. The simplest of these are the anti de Sitter (AdS) and the de Sitter (dS) spaces. Once again, the Klein-Gordon equation in such a background is solvable, and so all computations can be done in full detail. AdS is analogous to a magnetic background, while dS is analogous to an electric background. Analysis of the Bogoliubov transformation between certain ``in'' and ``out'' vacuum states indicates non-perturbative particle production in dS backgrounds \cite{candelas,lapedes,mottola,allen2,bousso}, but not in AdS backgrounds. Furthermore, in the de Sitter space, particle production  occurs only in even dimensions, not in odd dimensions  \cite{bousso}. The propagators and effective Lagrangian densities for dS and AdS have been computed in many different ways. Here we first briefly review these results, expressing them in compact forms suitable for comparing and contrasting with the effective Lagrangian density of Euler-Heisenberg for constant electromagnetic backgrounds described in the previous section. We then interpret these results in terms of large-order perturbation theory and Borel summation. 

A convenient route to the effective Lagrangian density is to consider the coincident limit of the Feynman propagator $G^{(F)}$, which is related to the effective Lagrangian density as :
\bea
\frac{\partial {\mathcal L}}{\partial m^2} = \frac{i}{2} \, G^{(F)}(x,x)\quad.
\label{lag}
\eea
Integrating this relation over $m^{2}$ then determines the effective Lagrangian density \cite{candelas,dowker,burgess,kamela}. Exactly equivalent results are obtained from mode expansions or using zeta functions \cite{dowker,isham,camporesi,dowker-sphere}.
The  Feynman propagators  for massive scalars in $AdS_d$ \cite{isham,burgess} and $dS_d$  \cite{candelas} have the forms:
\bea
-i G^{(F)}_{AdS_d}(x, x)&=&\frac{1}{K}\left(\frac{K}{4\pi}\right)^{d/2} \frac{\Gamma\left[1-\frac{d}{2}\right]\Gamma\left[\frac{d-1}{2}+\sqrt{\frac{m^2}{K}+\left(\frac{d-1}{2}\right)^2}\right]}{\Gamma\left[1-\frac{d-1}{2}+\sqrt{\frac{m^2}{K}+\left(\frac{d-1}{2}\right)^2}\right]}
\label{adsg}
\eea
\bea
-i G^{(F)}_{dS_d}(x, x)
= \frac{1}{K}\left(\frac{K}{4\pi}\right)^{d/2} 
\frac{\Gamma\left[1-\frac{d}{2}\right] 
\Gamma\left[\frac{d-1}{2}+i\sqrt{\frac{m^2}{K}-\left(\frac{d-1}{2}\right)^2}\right]  
\Gamma\left[\frac{d-1}{2}-i\sqrt{\frac{m^2}{K}-\left(\frac{d-1}{2}\right)^2}\right]}
{\Gamma\left[\frac{1}{2}+i\sqrt{\frac{m^2}{K}-\left(\frac{d-1}{2}\right)^2}\right]
\Gamma\left[\frac{1}{2}-i\sqrt{\frac{m^2}{K}-\left(\frac{d-1}{2}\right)^2}\right]}\nn\\
\label{dsg}
\eea
where $R=\pm d(d-1) K$ is the Ricci scalar, with $K$ positive. In our conventions, the curvature scalar is positive in anti de Sitter space and negative in de Sitter space.
 We expand about various integer values of the dimension $d$, and then integrate over $m^2$, to derive the effective Lagrangian density. For dimensions $d=2, 3, 4$, this coincides with the renormalized effective Lagrangian density \cite{candelas,dowker,burgess,kamela}, and in higher dimensions this procedure can be taken as a definition of the corresponding effective Lagrangian density. It is clear already from \eqn{adsg} and \eqn{dsg} that odd and even dimensions are very different.

Given the effective Lagrangian density, we can expand in a weak-field perturbative expansion, the gravitational analogues of \eqn{bexp} and \eqn{eexp} :
\bea
{\mathcal L}_{AdS_d}(K)&\sim& \left(\frac{m^2}{4\pi}\right)^{d/2} \sum_n a^{(AdS_d)}_n \left(\frac{K}{m^2}\right)^n \quad ,
\label{ads-pert}\\
{\mathcal L}_{dS_d}(K)&\sim& \left(\frac{m^2}{4\pi}\right)^{d/2}  \sum_n  a^{(dS_d)}_n \left(\frac{K}{m^2}\right)^n \quad .
\label{ds-pert}
 \eea
Notice that  in this weak-field expansion all powers  of the gravitational curvature $K$ appear, not just even powers as in the electromagnetic case.

\subsection{Odd Dimensions}
\label{odd-sec}

For $d$ odd (and $d\geq 3$), the effective Lagrangian densities  obtained from \eqn{lag}-\eqn{ds-pert} are:
\bea
{\mathcal L}_{AdS_d}(K)= \frac{(-1)^{\frac{d+1}{2}} \pi}{\Gamma\left(\frac{d}{2}\right)}\, \left(\frac{m^2}{4\pi}\right)^{d/2}\,
 \int_0^{\sqrt{1+\left(\frac{d-1}{2}\right)^2\frac{K}{m^2}}} dy\, y^2 \prod_{j=1}^{\frac{d-3}{2}}\left(y^2-\frac{K}{m^2} j^2\right)
 \label{ads-odd}
 \eea
 \bea
{\mathcal L}_{dS_d}(K)= \frac{(-1)^{\frac{d+1}{2}} \pi}{\Gamma\left(\frac{d}{2}\right)}\, \left(\frac{m^2}{4\pi}\right)^{d/2}\,
 \int_0^{\sqrt{1-\left(\frac{d-1}{2}\right)^2\frac{K}{m^2}}} dy\, y^2 \prod_{j=1}^{\frac{d-3}{2}}\left(y^2+\frac{K}{m^2} j^2\right) \coth\left(\pi \sqrt{\frac{m^2}{K}}\, y \right)\nn\\
 \label{ds-odd}
 \eea
It is clear that each of these integral representations can be expanded in a {\it convergent} perturbative weak-field expansion of the form in \eqn{ads-pert}-\eqn{ds-pert}, in powers of the curvature $K$. Furthermore, the coefficients of such an expansion are related by the replacement $K\to -K$ [compare with \eqn{ebalt}]:
\bea
a^{(AdS_d)}_n=(-1)^n a^{(dS_d)}_n \quad .
\label{odd-sign}
 \eea
Note that in the perturbative weak curvature limit, the coth factor in \eqn{ds-odd} reduces to unity. Since these expansions are convergent, there is no issue with Borel summation,  nor any indication of a non-perturbative imaginary contribution to the effective Lagrangian density. This is consistent with the absence of particle production in AdS or dS space in odd dimensions \cite{bousso}.

\subsection{Even Dimensions}
\label{even-sec}

In even dimensions, the situation is very different. For $AdS_d$, with $d$ even, one finds
\bea
{\mathcal L}_{AdS_d}(K)&=& \frac{2(-1)^{\frac{d}{2}+1}}{\Gamma\left(\frac{d}{2}\right)}\, \left(\frac{K}{4\pi}\right)^{d/2}\left\{ {\mathcal P}_d\left(\sqrt{\frac{m^2}{K}+\left(\frac{d-1}{2}\right)^2}\right)+\right.\nn\\
&&\left. \hskip -3cm \int_0^{\sqrt{\frac{m^2}{K}+\left(\frac{d-1}{2}\right)^2}} dx\, x \left[\prod_{j=0}^{\frac{d-4}{2}}\left(x^2-\left(j+\frac{1}{2}\right)^2\right)\right]\, \left(\psi\left(\frac{d-1}{2}+x\right)+\frac{1}{2} \ln \left(\frac{K}{4\pi\mu^2}\right)\right) \right\}\quad , 
\nn\\
 \label{ads-even}
 \eea
 where ${\mathcal P}_d$ is a polynomial of order $d$, whose specific form is known, but (being  a polynomial) is not important for a discussion of the divergent large order behavior of the weak curvature expansion of the effective Lagrangian density. When $d=2$, the product factor inside the integral in \eqn{ads-even} is absent. The scale $\mu^2$ is introduced by dimensional regularization, and in 4 dimensions it can be used to define a renormalized effective Lagrangian density \cite{burgess}. Using the identity \eqn{magic} derived in the appendix, this effective Lagrangian density can be expressed in a compact form in terms of  multiple gamma functions \cite{barnes}:
\bea
{\mathcal L}_{AdS_d}(K)&=&\frac{(-1)^{\frac{d}{2}} \Gamma\left(\frac{d+1}{2}\right) K^{d/2}}{2\pi^{(d+1)/2}}\, \left\{   \ln\left(\frac{\Gamma_d^2\left(\frac{d-1}{2}+\sqrt{\frac{m^2}{K}+\left(\frac{d-1}{2}\right)^2}\right)}{\Gamma_{d-1}\left(\frac{d-1}{2}+\sqrt{\frac{m^2}{K}+\left(\frac{d-1}{2}\right)^2}\right)}\right) +\right.\nn\\
&&\left. \hskip -2cm \tilde{{\mathcal P}}_d\left(\sqrt{\frac{m^2}{K}+\left(\frac{d-1}{2}\right)^2}\right)+\ln \left(\frac{K}{4\pi \mu^2}\right) \tilde{{\mathcal R}}_d\left(\sqrt{\frac{m^2}{K}+\left(\frac{d-1}{2}\right)^2}\right) \right\}
\label{ads-even-gamma}
\eea
Here $\tilde{{\mathcal P}}_d$ and $\tilde{{\mathcal R}}_d$ are polynomials of order $d$, whose explicit forms are not significant for our discussion. For $d=4$, the $AdS_4$ effective Lagrangian density was expressed in \cite{kamela} in terms of the multiple gamma functions $\Gamma_4$, $\Gamma_3$ and $\Gamma_2$, but using property \eqn{mg1} of the multiple gamma functions (see appendix), one can in general reduce this to the even more compact form \eqn{ads-even-gamma} in terms of just $\Gamma_d$ and $\Gamma_{d-1}$. 
 
 The analogous expressions for $dS_d$ are 
 \bea
{\mathcal L}_{dS_d}(K)&=& \frac{2(-1)^{\frac{d}{2}+1}}{\Gamma\left(\frac{d}{2}\right)}\, \left(\frac{K}{4\pi}\right)^{d/2}\left\{ {\mathcal P}_d\left(\sqrt{\frac{m^2}{K}-\left(\frac{d-1}{2}\right)^2}\right)+\right.\nn\\
&&\left. \hskip -3cm \int_0^{\sqrt{\frac{m^2}{K}-\left(\frac{d-1}{2}\right)^2}} dx\, x \left[\prod_{j=0}^{\frac{d-4}{2}}\left(x^2+\left(j+\frac{1}{2}\right)^2\right)\right]\, \left({\rm Re}\left[\psi\left(\frac{d-1}{2}+i x\right)\right]+\frac{1}{2}\ln \left(\frac{K}{4\pi\mu^2}\right)\right) \right\}
\nn\\
 \label{ds-even}
 \eea
where, as before, the product factor inside the integral is absent for $d=2$. Using the identity \eqn{magic} derived in the appendix, this can also be expressed in terms of multiple gamma functions:
\bea
{\mathcal L}_{dS_d}(K)&=& -\frac{1}{2}\frac{\Gamma\left(\frac{d+1}{2}\right) K^{d/2}}{2\pi^{(d+1)/2}}  \,  \left\{\tilde{{\mathcal P}}_d\left(\sqrt{\frac{m^2}{K}-\left(\frac{d-1}{2}\right)^2}\right)\right.\nn\\
&&\left. + \ln \left(\frac{K}{4\pi \mu^2}\right) \tilde{{\mathcal R}}_d\left(\sqrt{\frac{m^2}{K}-\left(\frac{d-1}{2}\right)^2}\right) \right. \nn\\ 
&&\left. \hskip -3cm +  \ln\left(\frac{\Gamma_d^2\left(\frac{d-1}{2}+i\sqrt{\frac{m^2}{K}-\left(\frac{d-1}{2}\right)^2}\right) \Gamma_d^2\left(\frac{d-1}{2}-i\sqrt{\frac{m^2}{K}-\left(\frac{d-1}{2}\right)^2}\right)}{\Gamma_{d-1}\left(\frac{d-1}{2}+i\sqrt{\frac{m^2}{K}-\left(\frac{d-1}{2}\right)^2}\right)\Gamma_{d-1}\left(\frac{d-1}{2}-i\sqrt{\frac{m^2}{K}-\left(\frac{d-1}{2}\right)^2}\right)}\right)
\right\}
 \label{ds-even-gamma}
 \eea
We recognize the prefactor as $-\frac{1}{2}$ times the inverse of the volume of $dS_d$. This agrees with the compact form for the log determinant of the Klein-Gordon operator, in terms of $\Gamma_d$ and $\Gamma_{d-1}$, first found using a zeta function approach by Voros \cite{voros} for $dS_2$, and generalized to $dS_n$ by Quine {\it et al} \cite{quine}. In the zeta function approach it also follows from the factorization of the sphere problem into two hemisphere problems \cite{dowker-sphere}, with Dirichlet and Neumann boundary conditions contributing, respectively, the multiple gamma terms:
\bea
{\rm Dirichlet:}&& \ln\left(\frac{\Gamma_d\left(\frac{d-1}{2}+i\sqrt{\frac{m^2}{K}-\left(\frac{d-1}{2}\right)^2}\right) \Gamma_d\left(\frac{d-1}{2}-i\sqrt{\frac{m^2}{K}-\left(\frac{d-1}{2}\right)^2}\right)}{\Gamma_{d-1}\left(\frac{d-1}{2}+i\sqrt{\frac{m^2}{K}-\left(\frac{d-1}{2}\right)^2}\right)\Gamma_{d-1}\left(\frac{d-1}{2}-i\sqrt{\frac{m^2}{K}-\left(\frac{d-1}{2}\right)^2}\right)}\right)
\nn\\
{\rm Neumann:}&& \ln\left(\Gamma_d\left(\frac{d-1}{2}+i\sqrt{\frac{m^2}{K}-\left(\frac{d-1}{2}\right)^2}\right) \Gamma_d\left(\frac{d-1}{2}-i\sqrt{\frac{m^2}{K}-\left(\frac{d-1}{2}\right)^2}\right)\right)
\nn\\
\eea

The perturbative weak-curvature expansions \eqn{ads-pert}-\eqn{ds-pert} of these $AdS_d$ and $dS_d$ effective Lagrangian densities can be derived using the asymptotic expansion of the digamma function \cite{as}
\bea
\psi(z)\sim \ln z -\frac{1}{2z}-\sum_{n=1}^\infty \frac{{\mathcal B}_{2n}}{2\, n\, z^{2n}}\quad , \quad z\to\infty\quad, \quad |{\rm arg}(z)|<\pi \quad ,
\label{psi}
\eea
or, equivalently, using the known asymptotic expansions of the multiple gamma functions \cite{adamchik}. It is clear from the form of the effective Lagrangian densities \eqn{ads-even} and \eqn{ds-even} that these perturbative weak-field expansion coefficients are related by
\bea
a_n^{(AdS_d)}=(-1)^n a_n^{(dS_d)}\quad ,
\label{even-sign}
\eea
just as in the odd dimensional case. However, in contrast to the odd-dimensional case, the even-dimensional weak-curvature expansions are {\it divergent} series. The {\it leading} large order behavior of the expansion coefficients can be extracted using \eqn{psi}:
\bea
a_n^{(AdS_d)}&\sim& \frac{{\mathcal B}_{2n+d}}{n(2n+d)}\sim 2(-1)^n \frac{\Gamma(2n+d-1)}{(2\pi)^{2n+d}}\\
a_n^{(dS_d)}&\sim& (-1)^n\frac{{\mathcal B}_{2n+d}}{n(2n+d)}\sim 2\, \frac{\Gamma(2n+d-1)}{(2\pi)^{2n+d}} \quad .
\label{grav-growth}
\eea
Thus, for $AdS_d$,  with $d$ even, the perturbative weak curvature expansion is a divergent alternating series, analogous to the situation \eqn{bexp} for a charged scalar field in a magnetic background field. This series is Borel summable, and real. On the other hand, for $dS_d$,  with $d$ even, the perturbative weak field expansion is a divergent {\it non-alternating} series, analogous to the situation \eqn{eexp} for a charged scalar in an electric background field. This series is not Borel summable. Nevertheless, the effective Lagrangian density is actually real, as is manifest from the expressions  \eqn{ds-even} and \eqn{ds-even-gamma}. (Incidentally, this can also be seen from the fact that the coincident limit of the Feynman propagator in \eqn{dsg} is manifestly real.)

However, suppose one only knew the perturbative weak-field expansion coefficients $a_n^{(dS_d)}$ of the effective Lagrangian density; or (more likely) only the leading large-order behavior \eqn{grav-growth} of these coefficients. Then, in the absence of further information, one would be tempted, by analogy with the electric field case discussed in Section \ref{gauge-sec}, to apply the Borel dispersion relation rule \eqn{imag-borel}, and deduce an imaginary part :
\bea
{\rm Im} \left({\mathcal L}_{dS_d}(K)\right) \sim e^{-2\pi m/\sqrt{K}}\qquad ({\rm incorrect})
\label{wrong}
\eea
The fact that the exponent depends linearly on $m$, rather than quadratically as in the electric field case \eqn{e-imag}, can be traced from \eqn{imag-borel} to the fact that the expansion is in powers of $\frac{K}{m^2}$, rather than in powers of $\left(\frac{E}{m^2}\right)^2$. However, \eqn{wrong} is not correct, as the full $dS_d$ effective Lagrangian density \eqn{ds-even} is manifestly real. Interestingly, though, the imaginary part \eqn{wrong} agrees with the particle production rate predicted by the Bogoliubov transformation analysis \cite{lapedes,mottola,allen2,bousso}. This puzzle of a real effective Lagrangian density in the de Sitter case, even though one might expect a gravitational analogue of Schwinger particle production, was pointed out long ago for $d=4$ by Candelas {\it et al} \cite{candelas} and Dowker {\it et al} \cite{dowker}, before the Bogoliubov transformation analyses had computed the rate of particle production \cite{lapedes,mottola,allen2,bousso}. Here we have re-phrased this in terms of Borel summation and the relation between the perturbative weak-field expansion and the full non-perturbative result.

\section{Resolution}

The resolution of this puzzle lies in the observation that while the weak-curvature perturbative expansion coefficients in the anti de Sitter and de Sitter backgrounds are related by the simple change of sign of the curvature $K\to -K$ as in \eqn{even-sign}, this is not true of the full (non-perturbative) Feynman propagators, and therefore of the full effective Lagrangian densities. Specifically,  the continuation $K\to -K$ does not map the de Sitter and anti de Sitter Feynman propagators onto one another \cite{isham,burgess,camporesi}. Indeed, using the reflection formula for the gamma function, we can express the $AdS_d$ coincident propagator \eqn{adsg} as
\bea
-i G^{(F)}_{AdS_d}(x, x)&=& \frac{1}{K}\left(\frac{K}{4\pi}\right)^{d/2} 
\frac{\Gamma\left[1-\frac{d}{2}\right] 
\Gamma\left[\frac{d-1}{2}+\sqrt{\frac{m^2}{K}+\left(\frac{d-1}{2}\right)^2}\right]  
\Gamma\left[\frac{d-1}{2}-\sqrt{\frac{m^2}{K}+\left(\frac{d-1}{2}\right)^2}\right]
}
{\Gamma\left[\frac{1}{2}+\sqrt{\frac{m^2}{K}+\left(\frac{d-1}{2}\right)^2}\right]
\Gamma\left[\frac{1}{2}-\sqrt{\frac{m^2}{K}+\left(\frac{d-1}{2}\right)^2}\right]
}\nn\\
&&\hskip -1cm \times \left[\sin \left(\pi \left(\frac{d-1}{2}\right)\right) -\tan\left(\pi \sqrt{\frac{m^2}{K}+\left(\frac{d-1}{2}\right)^2}\right)\, \cos \left(\pi \left(\frac{d-1}{2}\right) \right) \right]
\label{res}
\eea
The $\Gamma$ terms on the first line of \eqn{res} do continue into the $dS_d$ case \eqn{dsg} under $K\to -K$, but the factor on the second line does not, and is the source of the difference. Under $K\to -K$, the extra term $\tanh\left(\pi \sqrt{\frac{m^2}{K}-\left(\frac{d-1}{2}\right)^2}\right)$ is non-perturbative as far as a weak-curvature expansion is concerned, so this difference is not seen in the perturbative weak-curvature expansions \eqn{ads-pert}-\eqn{ds-pert}.

By contrast, in the electromagnetic case, under the continuation $B^2\to -E^2$, the Feynman propagators do continue into one another, with the appropriate boundary conditions for the magnetic and electric background fields, respectively \cite{nikishov,brezin,popov2}. This is true of the full background-field propagators, as well as of their weak-field expansions. But in the gravitational case, the continuation $K\to -K$ does not map the boundary conditions of the de Sitter and anti de Sitter cases into one another. For $dS_d$ and $AdS_d$, the propagator can be expressed in terms of hypergeometric functions, but the particular linear combination required to satisfy the Feynman boundary conditions is different in the two cases, and does not continue under $K\to -K$ \cite{isham,burgess,camporesi}. However, as is clear from \eqn{res}, this difference is not seen in the perturbative sector, but only in the non-perturbative sector.

From the viewpoint of Borel summation, the puzzle is resolved as follows. Consider the digamma function $\psi(x)=\frac{d}{dx}\ln\Gamma(x)$, which forms the kernel of the integral representations of the expressions for the effective Lagrangian densities in  \eqn{ads-even} and \eqn{ds-even}. The asymptotic expansion of the real part of $\psi(\frac{1}{2}+iy)$ is \cite{as}
\bea
{\rm Re}\left\{ \psi\left(\frac{1}{2}+i y\right)\right\}\sim \ln y-\sum_{n=1}^\infty \frac{(-1)^n \bar{\mathcal B}_{2n}}{2\,n\, y^{2n}}\quad ,
\label{realpsi}
\eea
for large real $y$.
The sum on the right hand side is a non-alternating divergent series for $y$ real, so the Borel formula \eqn{imag-borel} suggests it should have a non-perturbative imaginary part. But clearly the left hand side of \eqn{realpsi} is real! To resolve this apparent discrepancy, we apply the Borel relation \eqn{imag-borel} to the divergent sum and find
\bea
{\rm Im}\left\{ \sum_{n=1}^\infty \frac{(-1)^n \bar{\mathcal B}_{2n}}{2n y^{2n}} \right\}
=\frac{\pi}{2}\left(1-\tanh(\pi y)\right)\quad .
\label{psi-borel}
\eea
On the other hand, we also know that  \cite{as}
\bea
\psi\left(\frac{1}{2}+z\right)\sim \ln z-\sum_{n=1}^\infty \frac{\bar{\mathcal B}_{2n}}{2n z^{2n}}\quad , \quad z\to\infty\quad, \quad |{\rm arg}(z)|<\pi \quad ,
\label{full}
\eea
and furthermore (from the gamma function duplication formula) \cite{as}
\bea
{\rm Im}\left\{ \psi\left(\frac{1}{2}+i y\right)\right\}=\frac{\pi}{2}\tanh(\pi y) \quad .
\eea
Putting these together we see that in fact the real part \eqn{realpsi} has a non-perturbatively small (as $y\to\infty$) imaginary part
\bea
{\rm Re}\left\{ \psi\left(\frac{1}{2}+i y\right)\right\}\sim \ln y-\sum_{n=1}^\infty \frac{(-1)^n \bar{\mathcal B}_{2n}}{2n y^{2n}}+i\frac{\pi}{2}\left(1-\tanh(\pi y)\right) \quad ,
\label{real}
\eea
which exactly cancels the imaginary part \eqn{psi-borel} of the divergent sum, making the whole expression real, as it clearly must be. The imaginary part in \eqn{real} does not contribute to the perturbative asymptotic expansion at large real $y$. But, having ignored this term, it is inconsistent then to include the non-perturbative imaginary term \eqn{psi-borel} deduced from a Borel analysis of the divergent non-alternating perturbative expansion. Thus, if we only knew the perturbative expansion on the RHS of \eqn{realpsi}, we might erroneously deduce a non-perturbative imaginary part, which nevertheless is cancelled by the non-perturbative part in \eqn{real} once all non-perturbative contributions are included. 
Since the integral representations \eqn{ads-even}-\eqn{ds-even} of the $AdS_d$ and $dS_d$ effective Lagrangian densities are based on the $\psi$ function, this is exactly what happens when a na\"ive application of the Borel relation \eqn{imag-borel} suggests a non-perturbative imaginary part \eqn{wrong} for the even dimensional de Sitter effective Lagrangian density.

Perhaps even more interesting is the fact that this error actually gives the correct particle production rate, deduced from a Bogoliubov transformation argument. This is because the basic divergence of the weak-field expansion comes from the propagator's hypergeometric function at the coincident point. The Bogoliubov argument involves a different part of the propagator, that does have an imaginary part \cite{lapedes,mottola,allen2,bousso}, but it still has the same basic divergence property in its weak-field expansion. Ultimately, the physical reason for the apparent discrepancy is that $dS_d$ has a horizon, and so particle production is an observer dependent concept, as has been stressed by Gibbons and Hawking \cite{gibbons}. This means that 
particle production is not necessarily associated directly with an imaginary part of the effective Lagrangian density derived from the Feynman propagator.

\section{Conclusion}

In this paper we have analyzed the large-order behavior of the perturbative weak-field expansion of the effective Lagrangian density for a massive scalar field in de Sitter and anti de Sitter space. This is a gravitational analogue of the constant background field Euler-Heisenberg cases in QED. For AdS or dS in odd dimensions the effective Lagrangian density has a convergent perturbative expansion, consistent with the absence of non-perturbative particle production processes in odd dimensional AdS or dS space. In even dimensions the effective Lagrangian density has a divergent perturbative expansion. For AdS this divergent series is alternating and Borel summable, analogous to the case of a constant background magnetic field in the Euler-Heisenberg QED case. There is no  non-perturbative particle production. For even dimensional dS the divergent series is nonalternating, but in fact the effective Lagrangian density is real. Nevertheless, there is particle production, as found in a Bogoliubov analysis. This puzzle is resolved by a careful Borel analysis, and by noting that genuine non-perturbative information is needed concerning the Feynman propagators, and this information is not seen in the perturbative weak field expansion in even dimensional dS space. This shows that the connection between perturbative and non-perturbative physics is more subtle in the gravitational case than in the gauge background case. This may be of interest for more general gravitational effective actions \cite{avramidi}.

\section*{Acknowledgment}

This work was supported in part by US DOE under Grant numbers DE-FG-02-91ER40685 (AD) and  
DE-FG02-92ER40716 (GD).

\section{Appendix: Multiple Gamma Functions}
\label{mg-sec}

The multiple gamma functions $\Gamma_n(z)$ were introduced over a century ago by Barnes \cite{barnes}. They can be defined uniquely \cite{vigneras} by the conditions:
\bea
\Gamma_{n+1}(z+1)&=&\frac{\Gamma_{n+1}(z)}{\Gamma_n(z)}
\label{mg1} \\
\Gamma_1(z)&=&\Gamma(z)
\label{mg2}\\
\Gamma_n(1)&=&1
\label{mg3}\\
(-1)^{n+1}\frac{d^{n+1}}{dz^{n+1}} \ln \Gamma_n(z)&\geq& 0
\label{mg4}
\eea
Various integral representations and asymptotic expansions can be found in \cite{barnes,kamela,adamchik}. In some papers these functions are written in terms of $G_n(z)$ where 
$\Gamma_n(z)=\left[G_n(z)\right]^{(-1)^{n+1}}$. The most useful representation for our purposes can be derived from a result listed in \cite{kamela}:
\bea
\ln \Gamma_n(1+z)=\frac{(-1)^{n+1}}{(n-1)!}\int_0^z dx\, \left[\prod_{j=0}^{n-2}\left(x-j\right)\right]\,\psi(1+x)+{\mathcal Q}_n(z)\quad ,
\label{mg-integral}
\eea
where the ${\mathcal Q}_n(z)$ are known polynomials of degree $n$. From this it follows that with $n$ even:
\bea
\ln \left(\frac{\Gamma_n^2\left(\frac{n-1}{2}+z\right)}{\Gamma_{n-1}\left(\frac{n-1}{2}+z\right)} \right)&=&\frac{2}{(n-1)!}\int_0^z dx\, x \left[\prod_{j=0}^{\frac{n}{2}-2}\left(x^2-\left(j+\frac{1}{2}\right)^2\right)\right]\, \psi\left(\frac{n-1}{2}+x\right)\nn\\
&& +\tilde{\mathcal Q}_n(z)\quad .
\label{magic}
\eea
This is precisely the identity needed to connect the form  \eqn{ads-even} with \eqn{ads-even-gamma}, and \eqn{ds-even} with \eqn{ds-even-gamma}.

\end{document}